# Prolog Server Pages

Alin Suciu, Kalman Pusztai, Andrei Vancea
*Technical University of Cluj-Napoca*
*Department of Computer Science*
*{Alin.Suciu, Kalman.Pusztai}@cs.utcluj.ro, Andrei.Vancea@xanadu.ro*

## Abstract

*Prolog Server Pages (PSP) is a scripting language, based on Prolog, than can be embedded in HTML documents. To run PSP applications one needs a web server, a web browser and a PSP interpreter. The code is executed, by the interpreter, on the server-side (web server) and the output (together with the html code in witch the PSP code is embedded) is sent to the client-side (browser). The current implementation supports Apache Web Server. We implemented an Apache web server module that handles PSP files, and sends the result (an html document) to the client. PSP supports both GET and POST http requests. It also provides methods for working with http cookies. In the spirit of Open Source movement we chose not to implement from ground a Prolog compiler, but rather to use an existing product. We chose SWI-Prolog as the Prolog backend of our application. PSP is open source software, distributed under the LGPL license.*

## 1. Introduction

PSP is a scripting language, based on Prolog that can be embedded in HTML documents. Using PSP one can develop web-based applications having Prolog as the scripting language. The name itself (PSP) is inspired from similar technologies like ASP (Active Server Pages) with Visual Basic as scripting language and JSP (Java Server Pages) with Java as the scripting language.

Prolog is a logical programming language that is particularly suited for to programs that involve symbolic computations. For this reason it is a frequently used language in Artificial Intelligence [1, 5] where manipulation of symbols and inference about them is a common task. A Prolog program consists of a series of rules and facts. A program is run by presenting some query and seeing if this can be proved against the known rules and facts, using a sound and complete inference rule.

The PSP interpreter consists of an Apache Web Server module that handles PSP requests and communicates with SWI-Prolog, the Prolog backend of PSP.

## 2. Prolog Server Pages (PSP)

### 2.1. PSP scripts

A PSP script is usually located in one or more files having the ".psp" extension. A PSP file is basically an html document containing a series of pieces of PSP code embedded in the document. The PSP code is bracketed between "<?psp" and "?>". Such piece of code is called a chunk. Each chunk is passed individually to the PSP interpreter. The interpreter replaces each chunk with the output of its interpretation, provided by the Prolog system. The following example shows how the (in)famous "hello world" page can be generated using PSP in a simple and straightforward way.

```
<html>
<head>
<title>Hello World example</title>
</head>
<body>
<?psp
msg('Hello, World!').
?-msg(X), write(X).
?>
</body>
</html>
```

**Example 1. Hello World (hello.psp)**

After interpretation the following text is sent to the browser:

```
<html>
<head>
<title> PSP example </title>
</head>
<body>
Hello, World!
</body>
</html>
```

**Example 2. Result of the "hello.psp" script**

A chunk consists of a series of Prolog predicate definitions and one or more Prolog queries. They both end with a dot ("."). A query starts with a question mark and a dash ("?-"), the standard convention used in Prolog systems.

After reading the declaration of a predicate the interpreter immediately asserts it to the Prolog database.

A query is the action of asking the program of some information about the Prolog data base. When reading a query the interpreter tries to find one solution and stop after finding it or a after a failure.

Output Prolog predicates can be used to generate the HTML code. The alternative would be to use a dedicated package for generation of HTML code such as the one included in Pillow [2]. PSP redirects SWI-Prolog's standard stream to the HTTP client, so that the generation of dynamic HTML pages is very similar to the other technologies (ASP, JSP), by simply writing to the standard output stream.

## 2.2. HTML Forms

A HTML form is a section of a document containing normal content, markup, special elements called controls (checkboxes, radio buttons, menus, etc.), and labels on those controls. Users generally "complete" a form by modifying its controls (entering text, selecting menu items, etc.), before submitting the form to the Web Server. Each control is referred by its name. The information send by the user to the server using a certain control is called the value of the control. PSP provides methods for accessing the data received from the client. For each pair (control_name, control_value) the PSP interpreter asserts the fact:

    arg('control_name', 'control_value').

The following example shows a simple form that requests for the name and email of a user and calls the form handler "form_handler.psp".

```
<html>
<head>
<title> Form test </title>
</head>
<body>
<form action="form_handler.psp"
method="get">
<p>
<label for="firstname">First name: </label>
<input type="text" name="firstname">
<br>
<label for="lastname">Last name: </label>
<input type="text" name="lastname">
```

```
<br>
<label for="email">Email: </label>
<input type="text" name="email">
<br>
<input type="submit" value="Send">
<input type="reset">
</p>
</form>
</body>
</html>
```

**Example 3. Simple form (form.html)**

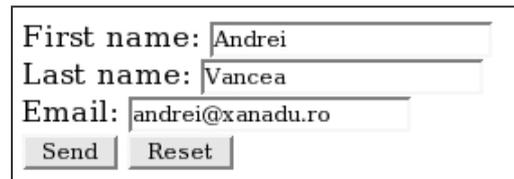

**Figure 1. "form.html" screenshot**

The form handler shown below executes a simple action, just printing the name and the email received from the form.

```
<html>
<head>
<title> Form handler </title>
</head>
<body>
<?psp
?-arg('firstname', FIRSTNAME),
write('First name : '),
write(FIRSTNAME),
write('<br>').
?-arg('lastname', LASTNAME),
write('Last name : '),
write(LASTNAME),
write('<br>').
?-arg('email', EMAIL),
write('Email :'),
write(EMAIL),
write('<br>').
?>
</body>
</html>
```

**Example 4. The form handler**

The result of the form handler is shown in the Figure 2 below.

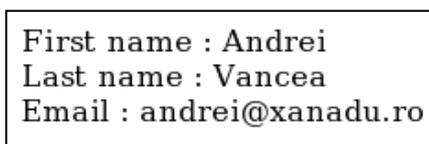

**Figure 2. "form_handler.psp" screenshot**

## 2.3 HTTP Cookies

Cookies are a mechanism used by the server-side to store and receive date to / from the client side. They a generally used for maintaining a persistent client / server connection. When the server response to an HTTP request it may also send a piece of data which the client will store. Any future requests from that client to the server will include all cookies received from the server. Cookies are transmitted into the HTTP header.

PSP offers two predicates for working with cookies: cookie/2 and setcookie/6.

The predicate cookie/2 has the syntax:

*cookie(+NAME, +VALUE)*

and retrieves a cookie already stored by the client. The arguments are:

NAME – the name of the cookie;

VALUE – the value of the cookie.

The predicate setcookie/6 has the syntax:

*setcookie(+NAME, +VALUE, +EXPIRES, +DOMAIN, +PATH, +SECURE)*

and writes a new cookie entry to the http response header. It has effects only if is called before any output predicate. The arguments are:

NAME – the name of the cookie;

VALUE – the value of the cookie;

EXPIRES - The expires attribute specifies a date string that defines the valid life time of that cookie. Once the expiration date has been reached, the cookie will no longer be stored or given out;

DOMAIN - When searching the cookie list for valid cookies, a comparison of the domain attributes of the cookie is made with the Internet domain name of the host from which the URL will be fetched. If there is a tail match, then the cookie will go through path matching to see if it should be sent. "Tail matching" means that domain attribute is matched against the tail of the fully qualified domain name of the host;

PATH - The path attribute is used to specify the subset of URLs in a domain for which the cookie is valid. If a cookie has already passed domain matching, then the pathname component of the URL is compared with the path attribute, and if there is a match, the cookie is considered valid and is sent along with the URL request;

SECURE - If a cookie is marked secure, it will only be transmitted if the communications channel with the host is a secure one. Currently this means that secure cookies will only be sent to HTTPS (HTTP over SSL) servers.

## 3. Implementation

In the spirit of Open Source movement we chose not to implement from ground a Prolog compiler, but rather to use an existing product. We chose SWI-Prolog as the Prolog backend of our application.

PSP can be considered an interface between SWI-Prolog and Apache Web Server. We developed an Apache Web Server module that handles PSP files, and sends the result (an HTML document) to the client (web browser).

This implementation was developed for Apache Web Server 2.0 and SWI-Prolog 5.0.

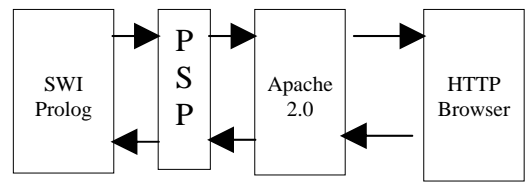

**Figure 3. Overview of PSP architecture**

From a development point of view, PSP consists of three parts:

a) Apache interface
b) SWI-Prolog interface
c) PSP Interpreter

The PSP interpreter was developed as an Apache module, called mod_psp, that runs in the Apache memory space. After loading, mod_psp registers itself as a handler for "text/psp" files. When a PSP request is made, the module opens the requested file and calls the interpreter routine. An important aspect about developing Apache modules is memory allocations. Because a module is loaded only once, when an Apache process starts, more that one requests can be handled in the same memory space. Memory allocated in the process of handling a request must be deallocated when the handling routine ends. Not doing so can cause severe memory leaks and make the system instable.

SWI-Prolog's powerful foreign interface made the development of PSP - Prolog interface relatively easy. SWI-Prolog provides C functions for asserting a predicate and for finding solutions of a Prolog query. It also allows defining Prolog predicates in C. Before the actual interpretation the PSP interpreter asserts an "arg(control_name, control_value)" fact for each HTTP argument. A "cookie" fact is asserted for each cookie included in the incoming HTTP header. We wrote a Prolog predicate ("setcookie") that writes a new cookie to the output HTTP header.

The PSP interpreter receives a chunk of Prolog code and returns the output resulting from the lateral effect of Prolog output predicates.

## 4. Conclusions

We developed an interface between SWI-Prolog, a robust Prolog implementation, and Apache HTTP Server, the most popular web server on the Internet.

PSP allows Prolog programmers to develop powerful web applications taking advantage of the advanced reasoning capabilities of Prolog.

PSP is intended to be an alternative to ASP and JSP for Prolog programmers as well as a complement to these technologies.

Thus, wherever Prolog seems more suitable for solving a specific task, PSP can be used together with other technologies (e.g. ASP, JSP).